\documentclass[aps,showpacs,nofootinbib,12pt]{revtex4}
\usepackage{amsmath}
\usepackage{amsfonts}
\usepackage{amssymb}
\usepackage{color}
\usepackage{graphicx}
\usepackage{mathrsfs}

\def\bc{\begin{center}}
\def\nno{\nonumber}
\def\ec{\end{center}}
\def\be{\begin{eqnarray}}
\def\ee{\end{eqnarray}}

\def\dS{dS}
\definecolor{dyellow}{rgb}{1.,0.8,.0}
\definecolor{myblue}{rgb}{.1,.1,.7}
\definecolor{dcyan}{rgb}{.0,.6,.6}
\definecolor{dmagenta}{rgb}{0.6,0.0,0.6}
\definecolor{brown}{rgb}{0.6,0.2,0.}
\definecolor{darkblue}{rgb}{.0,.0,0.5}
\definecolor{darkred}{rgb}{0.75,0.0,0.0}
\definecolor{orange}{rgb}{1.,.6,.0}
\definecolor{dorange}{rgb}{0.8,.4,.0}
\definecolor{darkgreen}{rgb}{0.0,0.6,0.0}
\definecolor{purple}{rgb}{.4,.0,.4}
\definecolor{lightgrey}{rgb}{0.7, 0.7, 0.7}
\definecolor{grey}{rgb}{0.4, 0.4, 0.4}

\def\La{\Lambda}

\def\dl{\delta}
\def\eps{\epsilon}
\def\ka{\kappa}
\def\la{\lambda}

\def\si{\sigma}

\def\d#1#2{\frac{\displaystyle #1}{\displaystyle #2}}

\newcommand\btd{\raise 2pt
\hbox{$\hat\bigtriangledown$}\hskip 1.5pt}
\newcommand\bt{\raise 2pt
\hbox{$\bigtriangledown$}\hskip 1.5pt}

\newcommand{\omits}[1]{}

\def\PRD{{Phys. Rev.}~{\bf D}}
\def\PRL{{Phys. Rev. Lett. }}
\def\PLA{{Phys. Lett.}~{\bf A}}

\begin{document}

\title{On Torsion-free Vacuum Solutions of the Model of de Sitter Gauge
Theory of Gravity}
\vskip 2cm

\author{Chao-Guang Huang$^{a,e}$\footnote{Email: huangcg@ihep.ac.cn}, Yu
Tian$^{b,e}$\footnote{Email: ytian@bit.edu.cn}, Xiaoning
Wu$^{c}$\footnote{Email: wuxn@amss.ac.cn} Han-Ying
Guo$^{d,e}$\footnote{Email: hyguo@ihep.ac.cn}}

\affiliation{\footnotesize $^a$ Institute of Theoretical Physics,
Chinese Academy of Sciences, Beijing 100049, China}

\affiliation{\footnotesize $^b$ Department of Physics, Beijing
Institute of Technology, Beijing 100081, China}

\affiliation{\footnotesize $^c$ Institute of Mathematics, Academy of
Mathematics and System Science, Chinese Academy of Sciences, Beijing
100080, China} \affiliation{\footnotesize $^d$ Institute of
Theoretical Physics, Chinese Academy of Sciences, Beijing 100080,
China}

\affiliation{\footnotesize $^e$ Kavli Institute for Theoretical
Physics China at the Chinese Academy of Sciences, Beijing, 100080,
China}

\vskip 3cm

\date{November 2007}

\newpage

\begin{abstract}
It is shown that all vacuum solutions of Einstein field equation
with a positive cosmological constant are the solutions of a model
of dS gauge theory of gravity. Therefore, the model is expected to
pass the observational tests on the scale of solar system and
explain the indirect evidence of gravitational wave from the binary
pulsars PSR1913+16.
\end{abstract}

\pacs{04.50.+h, 04.20.Jb}

\maketitle

The astronomical observations show that our universe is probably an
asymptotically de Sitter(dS) one \cite{SN,WMAP}. It arises the
interests on dS gauge theories of gravity. There is a model of the
dS gauge theory of gravity, which was first proposed in the 1970's
\cite{dSG, T77}. The model can be stimulated from dS invariant
special relativity \cite{dSSR, meetings, dSSR2} and the principle of
localization \cite{Guo2}, just like that the Poincar\'e gauge theory
of gravity may be stimulated from the Einstein special relativity
and the localization of Poincar\'e symmetry \cite{PGT}. The
principle of localization is that the full symmetry of the special
relativity as well as the laws of dynamics are both localized. The
gravitational action of the model takes the Yang-Mills form of
\cite{dSG,T77,Guo2}
\be
S_{\rm GYM}=\d 1 {4g^2} \int_{\cal M}d^4x e
{\bf Tr}_{\rm dS}({\cal F}_{\mu\nu}{\cal F}^{\mu\nu}),
\ee
where $e=\det(e^a_\mu)$ is the determinant of the tetrad $e^a_\mu$,
$g$ is a dimensionless coupling constant introduced as usual in the
gauge theory to describe the self-interaction of the gauge field,
\be \label{dScurv}
{\cal F}_{\mu\nu}=\left ({\cal F}^{AB}_{\ \ \ \mu\nu} \right )=
\left (\begin{array}{cc}
F^{ab}_{\ \ \mu\nu}+R^{-2} e^{ab}_{\ \ \mu\nu} & R^{-1}T^a_{\ \mu\nu}\\
-R^{-1}T^b_{\ \mu\nu} & 0\end{array} \right ) \omits{\in so(1,4)}
\ee
is the curvature of dS connection\footnote{The same \dS-connection with different
dynamics has also been explored in Ref. \cite{dSconnection}.}
\be
{\cal B}_{\mu}=\left ({\cal B}^{AB}_{\ \ \ \mu}\right)=\left(\begin{array}{cc}
B^{ab}_{\ \ \mu} & R^{-1}e^a_\mu \\
-R^{-1}e^b_\mu & 0
\end{array}
\right) \in so(1,4),
\ee
and ${\bf Tr}_{\rm dS}$ is the trace for the $so(1,4)$ indices $A,
B$. In Eq.(\ref{dScurv}), $F^{ab}_{~~\mu\nu}$ and $T^a_{~\mu\nu}$
are the curvature and torsion tensors of the Lorentz
connection $B^a_{\ b\mu}\in so(1,3)$, respectively, $R$ is
the dS radius, and $e_{ab}^{~\mu\nu}=e_a^\mu e_b^\nu-e_a^\nu
e_b^\mu$.
 In terms of $F^{ab}_{~~\,\mu\nu}$ and $T^{a}_{~\, \mu\nu}$, the
gravitational action can be rewritten as
\be
S_{\rm GYM}= -\int_{\cal M}d^4x e
\left[\frac{1}{4g^2}F^{ab}_{~~\mu\nu}F_{ab}^{~\mu\nu}-\chi(F-2\Lambda)
- \frac{\chi}{2} T^a_{~\mu\nu}T_a^{~\mu\nu} \right], \label{GYM}
\ee
where $F= \frac{1}{2} F^{ab}_{~\ \mu\nu}e_{ab}^{~\ \mu\nu}$ the scalar curvature, the same as
the action in the Einstein-Cartan theory, $\chi=1/({ 16}\pi G)$ is a dimensional coupling constant,
$\La = 3/R^2=3\chi g^2$ is the cosmological constant.

The gravitational field equations, obtained by the variation of the total action
\be\label{S_t}%
S_{\rm T}=S_{\rm GYM}+S_{\mathbf M}%
\ee%
with respect to $e^a_{~\mu},B^{ab}_{~~\mu}$, are
\be\label{Geq2}%
T_{a~~\, ||\nu}^{~\mu\nu } &-& F_{~a}^\mu+\frac{1}{2}F e_a^\mu -
\Lambda
e_a^\mu = 8\pi G( T_{{\rm M}a}^{~~\mu}+T_{{\rm G}a }^{~~\mu}), \\
\label{Geq2'}%
F_{ab~~\, ||\nu}^{~~\mu\nu} &=& R^{-2}(16\pi G S^{\quad \mu}_{{\rm M}ab}+S^{\quad \mu}_{{\rm G}ab}).%
\ee
Here, $S_{\rm M}$ is the action of the matter source with minimum
coupling, $||$ represents the covariant derivative using both
Christoffel symbol $\{^\mu_{\nu\ka}\}$ and connection
$B^a_{\ b\mu}$, $F_a^{~\mu}=-F_{ab}^{~~\mu\nu}e^b_\nu$,
\be
T_{{\rm M}a}^{~~\mu}&:=&-\d 1 e \d {\dl S_{\rm M}}{\dl e^a_\mu} \\
\label{emG}
T_{{\rm G}a}^{~~\mu}&:=&g^{-2} T_{{\rm F}a}^{~~\mu}+2\chi T_{{\rm
T}a}^{~~\mu} \ee
are the tetrad form of the stress-energy tensor for matter and
gravity, respectively, where
\be
\label{emF}
T_{{\rm F}a}^{~~\mu}&:=&-\frac{1}{4e} \frac{\delta} {\delta e^a_{\mu}}
\int d^4x e {\rm Tr}(F_{\nu\ka}F^{\nu\ka}) \nno \\
&=& e_{a}^\ka {\rm Tr}(F^{\mu \la}F_{\ka \la})-\frac{1}{4}e_a^\mu {\rm
Tr}(F^{\la \si} F_{\la \si}) \ee
is the tetrad form of the stress-energy tensor for curvature and
\be\label{emT}%
T_{{\rm T}a}^{~~\mu}&:=&-\frac{1}{4e} \frac{\delta} {\delta e^a_{\mu}}
\int d^4x e T^b_{\ \nu\ka}T_b^{\ \nu\ka}+T_{a\quad ||\nu}^{\ \mu\nu}\nno \\
&=& e_a^\ka T_b^{~\mu\la}T^{b}_{~\ka\la}-\frac{1}{4}e_a^\mu
T_b^{~\la\si}T^b_{~\la\si}
\ee%
the tetrad form of the stress-energy tensor for torsion, and
\be S_{{\rm M}ab}^{\quad \, \mu} =\d 1 {2\sqrt{-g}}\d {\dl S_{\rm M}
}{\dl B^{ab}_{\ \ \mu}} %
\ee%
and $S_{{\rm G}ab}^{\quad \,\mu }$ are spin currents for matter and
gravity, respectively. Especially, the spin current for gravity can
be divided into two parts,
\be\label{spG}%
S_{{\rm G}ab}^{\quad \, \mu}&=&S_{{\rm F}ab}^{\quad \, \mu}+2S_{{\rm
T}ab}^{\quad \,\mu}, \ee
where
\be%
S_{{\rm F}ab}^{\quad \, \mu}&:=&\d 1 {2\sqrt{-g}}\d {\dl }{\dl
B^{ab}_{\ \
\mu}}\int d^4 x \sqrt{-g}F =
-e^{~~\mu \nu}_{ab\ \ {||}\nu} = Y^\mu_{~\, \la\nu}
e_{ab}^{~~\la\nu}+Y ^\nu_{~\, \la\nu } e_{ab}^{~~\mu\la} \\
S_{{\rm T}ab}^{\quad \mu}&:=& \d 1 {2\sqrt{-g}}\d {\dl }{\dl B^{ab}_{\ \ \mu}}\d 1 4
\int d^4 x \sqrt{-g}T^c_{\ \nu\la}T_c^{\ \nu\la}
=T_{[a}^{~\mu\la}e_{b]\la}^{}
\ee%
are the spin current for curvature $F_{ab}^{\ \ \mu\nu}$ and
torsion $T_a^{~\mu\nu}$, respectively.

In Ref.\cite{wzc}, it is shown that all vacuum solutions of Einstein
field equation without cosmological constant are the solutions of
Eq.(\ref{Geq2}) and Eq.(\ref{Geq2'}) for the case of sourceless,
torsion-free, and vanishing cosmological constant. However, a
positive cosmological constant is vitally important for the dS gauge
theories of gravity. Without a positive cosmological constant, the
gravity should be a Poincar\'e or AdS one. Therefore, in order to
see whether the model of dS gauge theory of gravity can pass the
observational tests on the scale of solar system, it should be
important to explore if the vacuum solutions of Einstein field
equation with a positive cosmological constant do satisfy the
equations of the model.

The purpose of the present Note is to show that it is just the case.  Namely,
all vacuum
solutions of Einstein field equation with a positive cosmological
constant are the solutions of the torsion-free vacuum
equations of the model of dS gauge theory of gravity.

For the sourceless case, the torsion-free gravitational field
equations of the model reduce to
\be
 && {\cal R}_{~a}^\mu-\frac{1}{2}{\cal R} e_a^\mu +
\Lambda
e_a^\mu = -8\pi G( T_{{\rm M}a}^{~~\mu}+T_{{\rm R}a }^{~~\mu}), \label{ElEq}\\%
&&{\cal R}_{ab~~;\nu}^{~~\mu\nu} = 0, \label{YangEq}
\ee
where $T_{{\rm R}a}^{~~ \mu}=e_{a}^{\nu}T_{{\rm R}\ \nu}^{~\mu}$ the
tetrad form of the stress-energy tensor of Riemann curvature ${\cal
R}_{ab}^{~~\mu\nu}\omits{\in \mathfrak{so}(1,3)}$, and a semicolon
$;$ is the covariant derivative using both the Christoffel and Ricci
rotation coefficients. Eq. (\ref{ElEq}) is the Einstein-like
equation, while Eq.(\ref{YangEq}) is the Yang equation \cite{Yang}.
It can be shown \cite{wzc} that
\be\nno%
T_{{\rm R} \mu}^{~~\nu}
&=&{\cal R}_{ab \mu\la}{\cal R}^{ab\nu\la} - \frac{1}{4}\dl_\mu^\nu({\cal R}_{ab\la\ka}{\cal R}^{ab\la\ka})\nno \\
&=& \d 1 2 ({\cal R}_{\ka\si \mu\la}{\cal R}^{\ka\si \nu\la}+{\cal R}^*_{~\ka\si \mu\la} {\cal R}^{*\ka\si \nu\la}) \nno \\
&=&2C_{\la\mu}^{~~\ka\nu}{\cal R}^\la_\ka +\frac{{\cal R}}{3}({\cal R}_\mu^\nu -\frac{1}{4}{\cal R}\dl_\mu^\nu), \label{emR}
\ee%
where ${\cal R}_{\ka\si\mu\la}$ is the Riemann curvature tensor,
${\cal R}^*_{\ka\si\mu\la} =\frac 1 2 {\cal R}_{\ka\si \tau
\rho}\eps^{\tau \rho}_{~~\mu\la}$ is the right dual of the Riemann
curvature tensor, $C_{\la\mu\ka\nu}$ is the Weyl tensor. In the last
step in (\ref{emR}), the G\'eh\'eniau-Debever decomposition for the
Riemann curvature,
\be
{\cal R}_{\mu\nu\ka\la}=C_{\mu\nu\ka\la} + E_{\mu\nu\ka\la} + G_{\mu\nu\ka\la},
\ee
is used {\cite{GD}}, where
\be
E_{\mu\nu\ka\la} &=& \d 1 2 (g_{\mu\ka} S_{\nu\la}+g_{\nu\la} S_{\mu\ka}-g_{\mu\la} S_{\nu\ka}-g_{\nu\ka} S_{\mu\la}), \qquad\\
G_{\mu\nu\ka\la} &=& \d {\cal R} {12}(g_{\mu\ka}g_{\nu\la}-g_{\mu\la}g_{\nu\ka}), \\
S_{\mu\nu} &=& {\cal R}_{\mu\nu}-\d 1 4 {\cal R}g_{\mu\nu}.
\ee

On the other hand, the vacuum Einstein field equation with a
(positive) cosmological constant reads
\be
{\cal R}_{~\nu}^\mu-\frac{1}{2}{\cal R} \dl_\nu^\mu +
\Lambda \dl_\nu^\mu = 0.
\ee
It results in
\be
{\cal R}=4\La , \qquad \qquad {\cal R}_{~\nu}^\mu = \La \dl_\nu^\mu ,\label{Ricci}
\ee
and thus
\be
S_{\mu\nu}=0.
\ee
Since the Weyl tensor is totally traceless, the stress-energy tenor for Riemann curvature vanishes, {\it i.e.},
\be
T_{{\rm R} \mu}^{~~\nu} =0.
\ee
Therefore, all vacuum solutions of Einstein field equation with a
cosmological constant are solutions of Eq.(\ref{ElEq}). In addition,
the Bianchi identity
\be
{\cal R}^{\mu\nu}_{\ \ \la\si;\ka} +{\cal R}^{\mu\nu}_{\ \ \ka\la;\si}+{\cal R}^{\mu\nu}_{\ \ \si\ka;\la}=0
\ee
leads to
\be
0={\cal R}^{\mu\nu}_{\ \ \la\si;\nu} -{\cal R}^{\mu}_{\ \la;\si}+{\cal R}^{\mu}_{\ \si
;\la}={\cal R}^{\ \ \mu\nu}_{\la\si \ \ ;\nu}. \label{rYEq}
\ee
Namely, Yang equation (\ref{YangEq}) is also satisfied.
(The last step of Eq.(\ref{rYEq}) is valid because of Eq.(\ref{Ricci}).)

Therefore, we come to the conclusion that all vacuum solutions of
the Einstein field equation with a positive cosmological constant
are the torsion-free vacuum solutions of the model of dS gauge
theory of gravity. In particular, the dS, Schwarzschild-dS, and
Kerr-de Sitter metrics satisfy the Eqs.(\ref{Geq2}) and
(\ref{Geq2'}). Note that the Birkhoff theorem has been proved for
the gravitational theory (\ref{GYM}) without a cosmological constant
\cite{RauchNieh}. So, the model is expected to pass the
observational tests on the scale of solar system. In addition, the
model has the same metric waves as general relativity and thus is
expected to explain the indirect evidence of the existence of
gravitational wave from the observation data on the binary pulsar
PSR1913+16.

One might think that the above results are trivial because the Yang
equation does not appear at all if the torsion-free condition is
assumed in the action, in which case the tetrad and connection are
not independent. However, the torsion-free manifold is just the
specific situation of the the model.  There is no reason to set the
torsion to be zero before the variation.

In fact, it can be shown that all solutions of vacuum Einstein field
equation with a positive cosmological constant are also the vacuum,
torsion-free solutions of the field equations when the terms
\[
F_a^{\ \,\mu}F^a_{\ \,\mu}, \  e^a_\nu e^b_\mu F_a^{\ \,\mu}F_b^{\ \,\nu} , \
 e^{ab}_{\ \ \, \la\si}e^{cd}_{\ \ \, \mu\nu}F_{ab}^{\ \ \,\mu\nu}F_{cd}^{\ \ \,\la\si}
, \  e^b_{\si} e^{c}_{\mu}F_{ab\ \, \nu}^{\ \ \mu}F_{ac}^{\ \ \,\nu\si}
, \ e_a^\la e_b^{\si} T^a_{\ \mu\la}T^{b\mu}_{\ \ \si}
, \ e_a^\si e_b^{\mu} T^a_{\ \mu\la}T^{b\la}_{\ \ \si} \nno
\]
are added in the gravitational Lagrangian.  Obviously, the last two
terms have no contribution to the vacuum, torsion-free field
equations, while the middle two terms contribute the same as the
term $F_{ab}^{\ \ \mu\nu}F^{ab}_{\ \ \mu\nu}$ does thus only alter
the unimportant coefficients.  The first two terms add the term
$(R_{[a}^\mu e_{b]}^\nu)_{;\nu}$ in Yang equation and the
stress-energy tensor $R_{\mu \la} R^{\nu \la} -\frac 1 4 \dl_\mu^\nu
R_{\si \la} R^{\si \la}$ in Einstein equation. Both of them vanish
for the solutions of the vacuum Einstein equation with a positive
cosmological constant.

Obviously, the conclusion is still valid if the integral of the
second Chern form of the dS connection over the manifold is added in
the action. Finally, the similar discussions can be applied to the
AdS case as well.

\begin{acknowledgments}\vskip -4mm
We thank Z. Xu, B. Zhou and H.-Q. Zhang for useful discussions. This
work is supported by NSFC under Grant Nos. 90503002,
10605005, 10775140, 10705048, 10731080 and Knowledge Innovation
Funds of CAS (KJCX3-SYW-S03).
\end{acknowledgments}


\begin{thebibliography}{07}
\bibitem{SN}
A. G. Riess \textit{et al}.,
Astron. J. {\bf 116} (1998), 1009, astro-ph/9805201;
\omits{
\bibitem{SN2}
Supernovae Cosmology Project Collaboration,} S. Perlmutter \textit{et
al}., Astrophys. J. {\bf 517} (1999), 565, astro-ph/9812133.
\omits{
\bibitem{SN3}}
A. G. Riess \textit{et al}., Astrophys. J. {\bf 536} (2000), 62,
astro-ph/0001384.
\omits{
\bibitem{SN4}
Supernova Search Team Collaboration,} A. G. Riess \textit{et al}.,
Astrophys. J. {\bf 560} (2001), 49, astro-ph/0104455.

\bibitem{WMAP}
C. L. Bennett \textit{et al}., Astrophys. J. Suppl. {\bf 148} (2003),
1, astro-ph/0302207;
\omits{
\ddot{}\bibitem{WMAP2}
WMAP Collaboration,} D. N. Spergel \textit{et al}., Astrophys. J.
Suppl. {\bf 148} (2003), 175, astro-ph/0302209.

\bibitem{dSG}Y.-S. Wu, G.-D. Li and H.-Y. Guo, Kexue Tongbao (Chi. Sci. Bull.)
{\bf 19} (1974), 509; Y. An, S, Chen, Z.-L. Zou and H.-Y. Guo, {\it ibid}, 379;
H.-Y. Guo, {\it ibid} {\bf 21} (1976) 31;  Z. L. Zou,  et al,  Sci. Sinica {\bf
XXII} (1979), 628; M.-L. Yan, B.-H. Zhao and H.-Y. Guo,
\omits{Renormalization of gravitation field with torsion,}
Kexue Tongbao (Chi. Sci. Bull.) {\bf 24} (1979), 587; {\it Acta
Physica Sinica} {\bf 33} (1984), 1377; 1386 (all in Chinese). 


\bibitem{T77} P. K. Townsend, \PRD15 (1977), 2795; A. A. Tseytsin, \PRD26
(1982), 3327.



\bibitem{dSSR} K.-H. Look (Q.-K. Lu) 1970, Why the Minkowski metric must be
used? unpublished; K.-H. Look, C.-L. Tsou (Z.-L. Zou) and H.-Y. Kuo (H.-Y.
Guo),  Acta Phys. Sin. {\bf 23} (1974), 225;  Nature (Shanghai, Suppl.);
H.-Y. Guo,
Kexue Tongbao (Chinese Sci. Bull.) {\bf 22} (1977), 487 (all in Chinese).

\bibitem{meetings}\omits{A
theory of inertial motion in space-time of constant curvature,}H.-Y. Guo, in
{\it Proceedings of the 2nd Marcel Grossmann Meeting on General Relativity}, ed. by R. Ruffini,
(North-Holland, 1982), 801; \omits{The meaning of relativity in
spacetimes of constant curvature,} {\it Nucl. Phys. B} (Proc.
Suppl.) {\bf 6} (1989), 381; C.-G. Huang and H.-Y. Guo, in {\it Gravitation and Astrophysics ---
On the Occasion of the 90th Year of General Relativity,
Proceddings of the VII Asia-Pacific International Conference}, ed. by J. M. Nester, C.-M Chen,
and J.-P. Hsu (World Scientific, 2007, Singapore), 260.

\bibitem{dSSR2}
H.-Y. Guo, C.-G. Huang, Z. Xu and B. Zhou,\omits{On Beltrami model
of de Sitter spacetime, arXiv:} {\it Mod. Phys. Lett.} {\bf A19}
(2004), 1701, 
hep-th/0403013;
\PLA331 (2004), 1, hep-th/0403171;
H.-Y. Guo, C.-G. Huang, Z. Xu and B. Zhou, {\it Chin. Phys. Lett.}
{\bf 22} (2005), 2477, hep-th/0508094;
 H.-Y. Guo, C.-G. Huang,  Y. Tian, Z. Xu and B. Zhou,
 {\it Acta Phys. Sin.} {\bf 54} (2005), 2494 (in Chinese); 
 H.-Y. Guo, C.-G. Huang and B. Zhou, {\it Europhys. Lett.} {\bf 72} (2005),
 1045, hep-th/0404010.

\bibitem{Guo2}
H.-Y. Guo, C.-G. Huang, Y. Tian, H.-T. Wu, B. Zhou, Class. Quant.
Grav. {\bf 24} (2007) 4009; H.-Y. Guo, C.-G. Huang, Y. Tian, B.
Zhou, Front. Phys. China, {\bf 2} (2007), 358; H.-Y. Guo, {\it
Phys. Lett.} {\bf B653} (2007) 88; H.-Y. Guo, Special Relativity and
Theory of Gravity via Maximum Symmetry and Localization -- In Honor
of the 80th Birthday of Professor Qikeng Lu, arXiv:0707.3855.



\bibitem{PGT} T. W. B. Kibble, {\it J. Math. Phys.} {\bf 2} (1961),
212; F. W. Held, P. von der Heyde, G. D. Kerlick,
and
J. M.  Nester,  
 {\it Rev Mod. Phys.} {\bf 48} (1976), 393 and references therein.

\omits{\bibitem{Wise} D.K. Wise, MacDowell-Mansouri Gravity and Cartan
Geometry, arXiv: gr-qc/0611154.}
\bibitem{dSconnection}  S. W. MacDowell and F. Mansouri, \PRL\ {\bf 38} (1977), 739;
Erratum-ibid. {\bf 38} (1977), 1376;
K. S. Stelle and P. C. West, \PRD21 (1980), 1466;
F. Wilczek, \PRL\ {\bf 80} (1998), 4951;
L. Freidel and A. Starodubtsev, Quantum gravity in terms of topological
observables, arXiv: hep-th/0501191;
M. Leclerc,  Annals of Physics, {\bf 321} (2006), 708; E. Witten,
Three-Dimensional Gravity Reconsidered,
arXiv:0706.3359.

\omits{
\bibitem{gtg79}
 H.Y. Kuo,
 in {\it Proc.   2nd M. Grossmann Meet. on GR} (1979), ed. by R. Ruffini,
 North-Holland Publ. (1982) 475.}

\bibitem{Yang}
C. N. Yang, Phys. Rev. Lett. {\bf 33} (1974), 445.

\bibitem{wzc}Y.-S. Wu, Z.-L. Zou and S. Chen, {\it Kexue Tongbao (Chin. Sci.
Bull.)} {\bf 18} (1973), 119 (in Chinese).

\bibitem{GD} H. Stephani, D. Kramer, M. MacCallum, C. Hoenselaers, E.
Herlt, Exact Solutions of Einstein's Field Equations, (Cambridge
University Press, Cambridge, 1980).

\bibitem{RauchNieh} R. Rauch and N. T. Nieh, Phys. Rev. D{\bf 24} (1981), 2029.


\end{thebibliography}
\end{document}